\documentclass{aa}  

\usepackage{graphicx}
\usepackage{natbib}
\newcommand{\teff}{\mbox{$T_{\rm eff}$}}
\newcommand{\logg}{\mbox{$\log g$}}
\newcommand{\vsini}{\mbox{$v \sin i$}}

\newcommand{\kms}{\mbox{km\,s$^{-1}$}}

\newcommand{\degree}{\ensuremath{^\circ}}

\def\ms{\hbox{\,m\,s$^{-1}$}}         
\def\m2s2{\hbox{\,m$^{2}$\,s$^{-2}$}} 
\def\kms{\hbox{\,km\,s$^{-1}$}}       
\def\vsini{\hbox{$v$\,sin\,$i$}}      
\def\Msun{\hbox{$M_{\odot}$}}             
\def\Rsun{\hbox{$R_{\odot}$}}
\def\Mjup{\hbox{$\mathrm{M}_{\rm Jup}$}}
\def\Rjup{\hbox{$\mathrm{R}_{\rm Jup}$}}

\def \1s{$1\,\sigma$}

\def \t0{T$_0$}

\def\Rsun{\hbox{$R_{\odot}$}}

\usepackage{txfonts}
\begin{document}

   \title{Discovery of WASP-113b and WASP-114b, two inflated hot-Jupiters with contrasting densities }



\author{S. C.  C.~Barros \inst{1,2}
\and D. J. A..~Brown  \inst{\ref{war}}
\and G.~H\'ebrard \inst{4,5}
\and Y.~G{\'o}mez Maqueo Chew \inst{\ref{war},\ref{mex} }
\and D. R. ~Anderson \inst{\ref{keele}}
\and P. ~Boumis \inst{\ref{athe}}
\and L.~Delrez  \inst{\ref{liege}}
\and K. L.~Hay  \inst{\ref{stand}}
\and K. W. F. ~Lam \inst{\ref{war}}
\and J. ~Llama \inst{\ref{lowell}}
\and M. ~Lendl \inst{\ref{austria}, \ref{gen}}
\and J. ~McCormac \inst{\ref{war}}
\and B. Skiff \inst{\ref{lowell}}
\and B ~Smalley \inst{\ref{keele}}
\and O ~Turner \inst{\ref{keele}}
\and M. ~Vanhuysse \inst{\ref{over}}
\and D. J. ~Armstrong \inst{\ref{war}, \ref{belfast}}
\and I. ~Boisse \inst{2}
\and F. ~Bouchy \inst{2}
\and A. ~Collier Cameron \inst{\ref{stand}}
\and F. ~Faedi \inst{\ref{war}}
\and M. Gillon \inst{\ref{liege}}
\and C. ~Hellier  \inst{\ref{keele}}
\and E. ~Jehin \inst{\ref{liege}}
\and A. ~Liakos \inst{\ref{athe}}
\and J. ~Meaburn \inst{\ref{manch}}
\and H. P.~Osborn\inst{\ref{war}}
\and F. ~Pepe  \inst{\ref{gen}}
\and I. Plauchu-Frayn \inst{\ref{mex2}}
\and D. ~Pollacco \inst{\ref{war}}
\and D. ~Queloz \inst{\ref{gen},\ref{cam}}
\and J. ~Rey  \inst{\ref{gen}}
\and J. ~Spake \inst{\ref{war}}
\and D. S\'egransan  \inst{\ref{gen}}
\and A. H. M. ~Triaud \inst{\ref{tor1}, \ref{tor2},\ref{gen}, \ref{cam2}}
\and S. ~Udry  \inst{\ref{gen}}
\and S. R. ~Walker \inst{\ref{war}}
\and C. A.~Watson  \inst{\ref{belfast}}
\and R. G. ~West \inst{\ref{war}}
\and P. J. ~Wheatley \inst{\ref{war}} }

   \institute{Instituto de Astrof\'isica e Ci\^encias do Espa\c{c}o, Universidade do Porto, CAUP, Rua das Estrelas, PT4150-762 Porto, Portugal\\
\email{susana.barros@astro.up.pt}
\and Aix Marseille Universit\'e, CNRS, LAM (Laboratoire d'Astrophysique de Marseille) UMR 7326, 13388, Marseille, France 
\and Department of Physics, University of Warwick, Gibbet Hill Road, Coventry, CV4 7AL, UK \label{war}
\and Institut d'Astrophysique de Paris, UMR7095 CNRS, Universit\'e Pierre \& Marie Curie, 98bis boulevard Arago, 75014 Paris, France
\and Observatoire de Haute-Provence, Universit\'e d'Aix-Marseille \& CNRS, 04870 Saint Michel l'Observatoire, France
\and Instituto de Astronom\'ia, Universidad Nacional Aut\'onoma de M\'exico, Ciudad Universitaria, D.F. 04510, M\'exico \label{mex}
\and Astrophysics Group, Keele University, Staffordshire ST5 5BG, UK \label{keele}
\and Institute for Astronomy, Astrophysics, Space Applications and Remote Sensing, National Observatory of Athens, 15236 Penteli, Greece \label{athe}
\and Institut d'Astrophysique et de Geophysique, Universite de Liege,  Allee du 6 Aout, 17, Bat. B5C, Liege 1, Belgium \label{liege}
\and SUPA, School of Physics and Astronomy, University of St. Andrews, North Haugh, Fife KY16 9SS, UK \label{stand}
\and Lowell Observatory, 1400 W Mars Hill Rd. Flagstaff. AZ. 86001. USA \label{lowell}
\and Space Research Institute, Austrian Academy of Sciences, Schmiedlstr. 6, 8042 Graz, Austria \label{austria}
\and OverSky, 47 allee des Palanques, BP 12, 33127, Saint-Jean d’Illac, France \label{over}  
\and Jodrell Bank Centre for Astrophysics, University of Manchester, Oxford Rd., Manchester, UK. M13 9PL \label{manch}
\and Observatoire de Geneve, Universite de Geneve, 51 Chemin des Maillettes, 1290 Sauverny, Switzerland \label{gen}
\and Instituto de Astronom\'ia, Universidad Nacional Aut\'onoma de M\'exico, C.P. 22860, Ensenada, Baja California, M\'exico  \label{mex2}
\and Cavendish Laboratory, J J Thomson Avenue, Cambridge, CB3 0HE, UK \label{cam}
\and Centre for Planetary Sciences, University of Toronto at Scarborough, 1265 Military Trail, Toronto, ON, M1C 1A4, Canada \label{tor1}
\and Department of Astronomy \& Astrophysics, University of Toronto, Toronto, ON M5S 3H4, Canada \label{tor2}
\and Institute of Astronomy, Madingley Road, Cambridge, CB3 0HA, United Kingdom \label{cam2}
\and Astrophysics Research Centre, School of Mathematics \& Physics, Queen's University Belfast, University Road, Belfast, BT7 1NN, UK \label{belfast}
 }

\date{Received September 15, 1996; accepted March 16, 1997}

  \abstract
{}
   {We present the discovery and characterisation of the exoplanets WASP-113b and WASP-114b by the WASP survey, {\it SOPHIE} and {\it CORALIE}.}
   { The planetary nature of the systems was established by performing follow-up photometric and spectroscopic observations. The follow-up data were combined with the WASP-photometry and analysed with an MCMC code to obtain system parameters.}
   {The host stars WASP-113 and WASP-114 are very similar.  They are both early G-type stars with an effective temperature of $\sim5900\,$K, [Fe/H]$\sim 0.12$ and \logg\ $\sim 4.1$dex. However, WASP-113 is older than WASP-114. Although the planetary companions have similar radii, WASP-114b is almost 4 times heavier than WASP-113b.\ WASP-113b has a mass of $0.48\,$ \Mjup\ and an orbital period of $\sim 4.5\,$days; WASP-114b has a mass of $1.77\,$ \Mjup and an orbital period of $\sim 1.5\,$days. Both planets have inflated radii, in particular WASP-113 with a radius anomaly of  $\Re=0.35$. The high scale height of WASP-113b ($\sim 950$ km ) makes it a good target for follow-up atmospheric observations.}
   {}


\maketitle
%

\section{Introduction}

In the last few years there has been a huge increase in the number of transiting exoplanets known mainly due to the Kepler satellite  \citep{Borucki2010}.
Currently, 2600 transiting exoplanets are known and there are another thousand unconfirmed candidates.
  Transiting planet systems are especially valuable because their geometry enables us to
derive accurate planetary properties \citep{Charbonneau2000, Henry2000}. Time series
photometry during transit allows estimation of the orbital inclination and the relative radii of the
host star and planet. These can be combined with radial velocity measurements and stellar parameters
to derive the absolute planetary mass  \cite[e.g.][]{Barros2011a}. Hence, the bulk density of the
planet can be estimated with good accuracy, giving us insight into its composition \citep{Guillot2005,Fortney2007}, thus placing constraints on planetary structure and formation models.

Furthermore, follow-up observations of transiting planets gives further insight into their physical properties.
Transmission spectroscopy, which consists of measuring the
stellar light filtered through the planet's atmosphere during transit, provides information about
exoplanet atmospheres \citep{Charbonneau2002, Vidal-Madjar2003}. Moreover, observation of secondary
eclipses (i.e. occultations) offers the potential for directly measuring planetary emission spectra \cite[e.g.][]{Deming2005, Charbonneau2008, Grillmair2008}. However, currently these follow-up observations are only feasible for bright stars.

The number of transiting exoplanets around stars brighter than V=13 is just a few hundred. These bright host stars enable follow-up observations to  better characterise the systems.
Therefore, several second generation surveys are
being developed to target bright stars both ground based (NGTS, MASCARA and SPECULOOS) and space based: CHEOPS (ESA), PLATO (ESA) and TESS (NASA).
The new Kepler satellite mission K2 \citep{Howell2014} is making a significant scientific impact due to monitoring a large number of fields and significant number of bright stars at a photometric precision only slightly inferior to the original Kepler mission \cite[e.g.][]{Vanderburg2014,Barros2016}

In this paper, we report the discovery of WASP-113b and WASP-114b by {\it SOPHIE}, {\it CORALIE} and the WASP-project \citep{Pollacco2006} that is the leading ground-based transit survey having discovered $\sim 150$ exoplanets around stars brighter than V=13 mag.
The WASP project consists
of two robotic observatories: one in the Observatorio del Roque de los
Muchachos, La Palma, Canary Islands, Spain and the other in the South
African Astronomical Observatory of Sutherland, South Africa.

The host stars WASP-113 and WASP-114 are similar early G type stars. Both planets have radius $\sim50$\% higher than Jupiter, but WASP-113b is less than half Jupiter's mass while WASP-114b has almost twice the mass of Jupiter. Hence WASP-114b is 4.2 times more dense than WASP-113b.
We start by describing the photometric and spectroscopic observations of both systems in Section~2 and present the spectroscopic characterisation of the stars in Section~3. In Section~4 we describe the system analysis and present the results and we finish by discussing our results in Section~5.


\section{Observations}

\subsection{SuperWASP observations}

The SuperWASP-North observatory in La Palma consists of 8 cameras each
with a Canon 200-mm f/1.8 lens coupled to an Andor e2v $2048 \times
2048$ pixel back-illuminated CCD \citep{Pollacco2006}. This
configuration gives a pixel scale of 13.7\arcsec/pixel which
corresponds to a field of view of $7.8\times 7.8$ square degrees per
camera.

The field containing WASP-113 
($\alpha=$ 14:59:29.49$\, \delta=$ +46:57:36.4) was observed in the
period between 30 March 2011 and 30 June 2011 simultaneously by two out of the 8 cameras.
A total of 29942 good points were collected for WASP-113.

The field containing WASP-114 
($\alpha=$ 21:50:39.74$\, \delta=$ +10:27:46.9) was observed from July 2006 till November 2011. During this period 53673 observations of WASP-114 were collected. 

\begin{figure}
  \centering
  \includegraphics[width=\columnwidth]{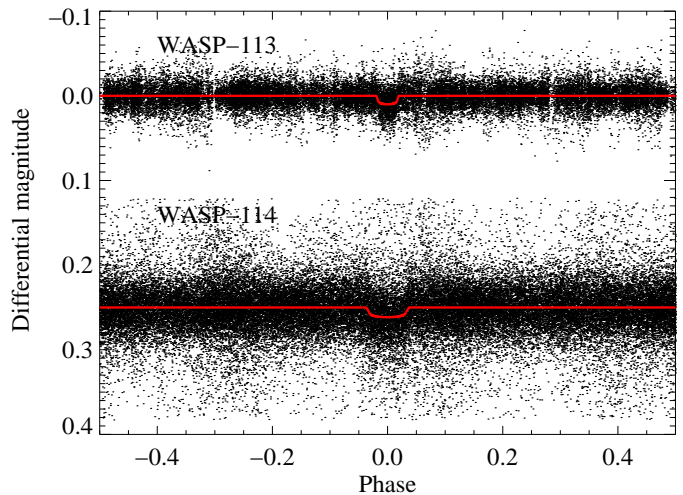}
  \caption{SuperWASP phase folded light curve for WASP-113 (top) and WASP-114 (bottom). The best transit model described in detail in section 4 is overplotted. The data of WASP-114 was displaced vertically for clarity.}
  \label{superwasplc}
\end{figure}

The light curves were analysed with the automatic WASP pipeline. The data are detrended using the algorithms SYSREM
\citep{Tamuz2005} and TFA \citep{Kovacs2005}. Subsequently a transit search is performed which is detailed in \citet{Cameron2006, Cameron2007}. Both stars were flagged as transit candidates after passing initial tests for false positives and were recommended for follow-up observations in March 2013 and June 2012 for WASP-113 and WASP-114 respectively.
The phase folded WASP light curves of WASP-113 and WASP-114 are shown in Figure~\ref{superwasplc}.

\subsection{Spectroscopic follow-up}

Radial velocity measurements of WASP-113 were taken with  {\it SOPHIE}
mounted on the 1.93m telescope of the Observatoire de Haute Provence
\citep{Perruchot2008,Bouchy2009}. A total of 20 measurements were obtained from the 21 April 2013 to the 16 of September 2013. 
For WASP-114 both {\it SOPHIE} and {\it CORALIE}  mounted on the 1.2m Swiss
Euler telescope in La Silla \citep{Baranne1996, Queloz2000, Pepe2002} were employed to obtain radial velocity measurements.  Between 13 July 2013 and 6 October 2013 18 measurements were obtained with  {\it SOPHIE} and 17 with {\it CORALIE}. The data were reduced with the {\it SOPHIE} and {\it CORALIE}
pipelines, respectively. The radial velocity errors account for the
photon noise.

\begin{figure*}
\centering
\includegraphics[width=0.45\textwidth]{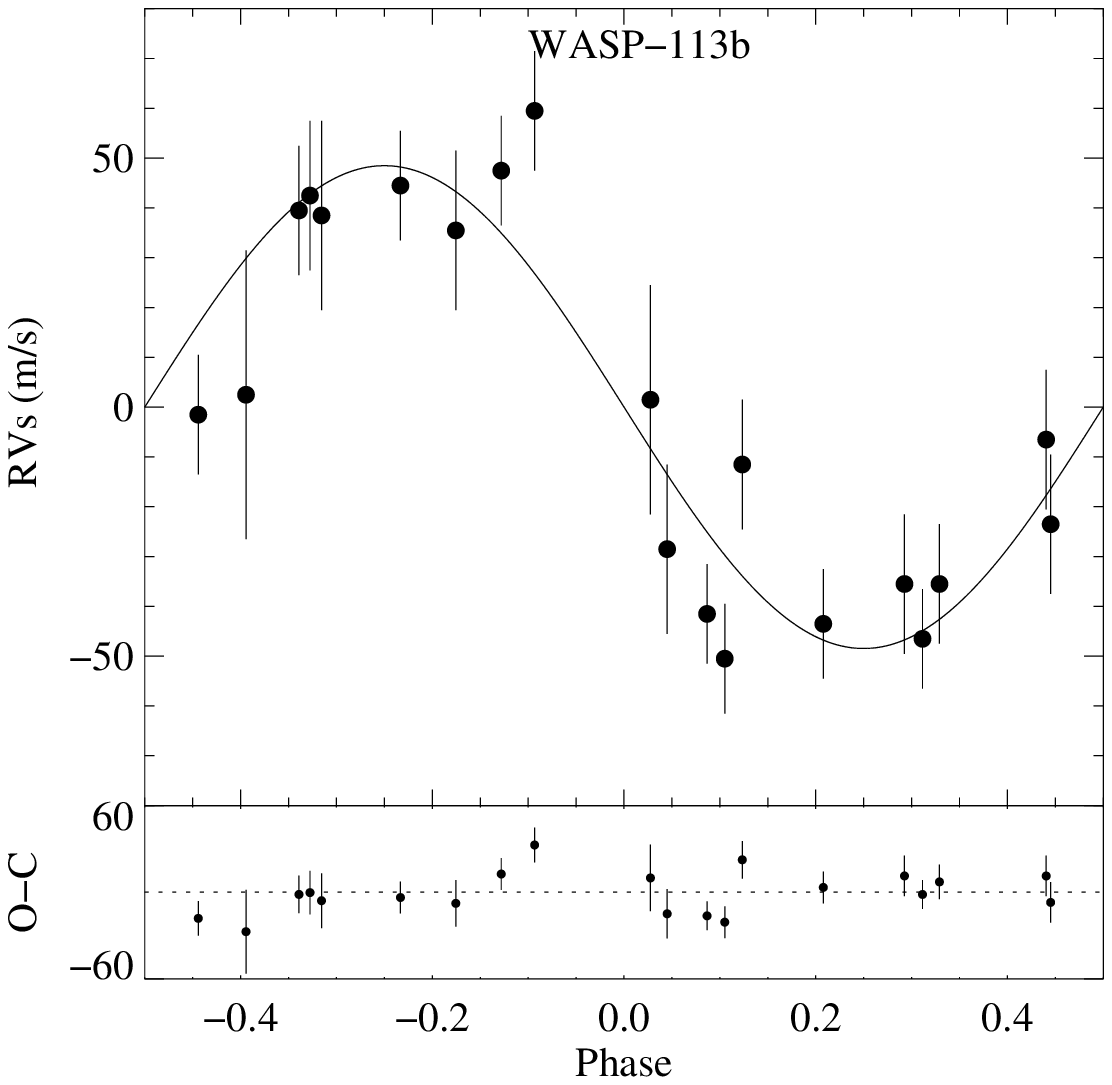}
\includegraphics[width=0.45\textwidth]{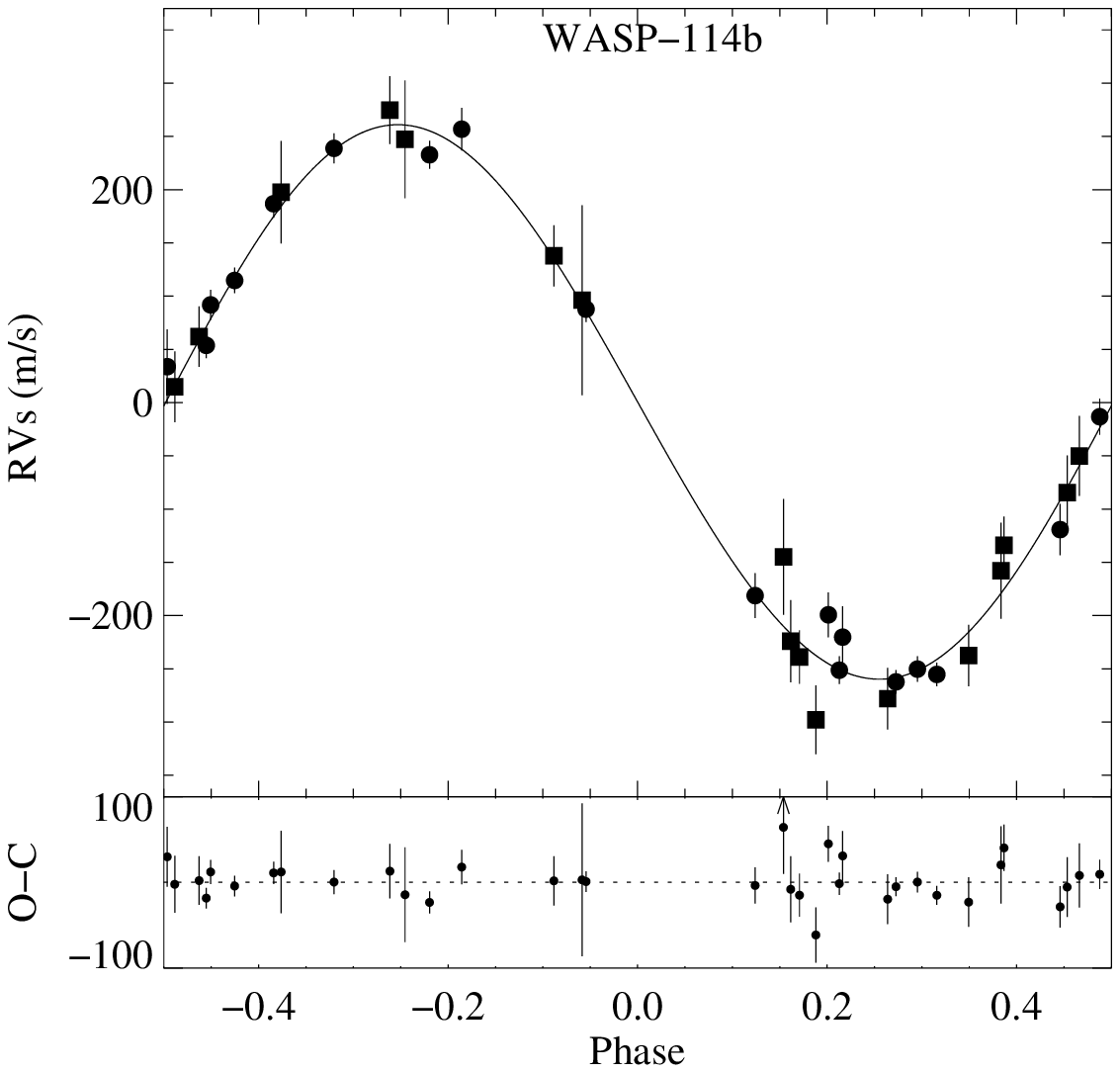}
\caption{ Phase folded radial velocities of WASP$-$113 (left) and WASP-114 (right). The observations taken with {\it SOPHIE} are plotted as circles and the ones taken with {\it CORALIE} as squares. The overplotted black curve is the most probable fit orbital model and the residuals from these fits are also given (bottom panel). For WASP-113b we impose a circular orbit while the eccentricity was left free in the model of WASP-114b. \label{rvs}}
\end{figure*}

The radial velocity measurements are given in
Table~\ref{rvobservations} for WASP-113 and in Table~\ref{rvobservations2} for WASP-114.
In Figure~\ref{rvs}, we show the phase folded radial velocities for WASP-113b in the top panel and WASP-114b in the bottom panel. 
The best fit Keplerian model described in Section~4 is superimposed on the data points and  we also show the residuals from the model, which show no long term trend. 

\begin{figure*}
\centering
\includegraphics[width=0.45\textwidth]{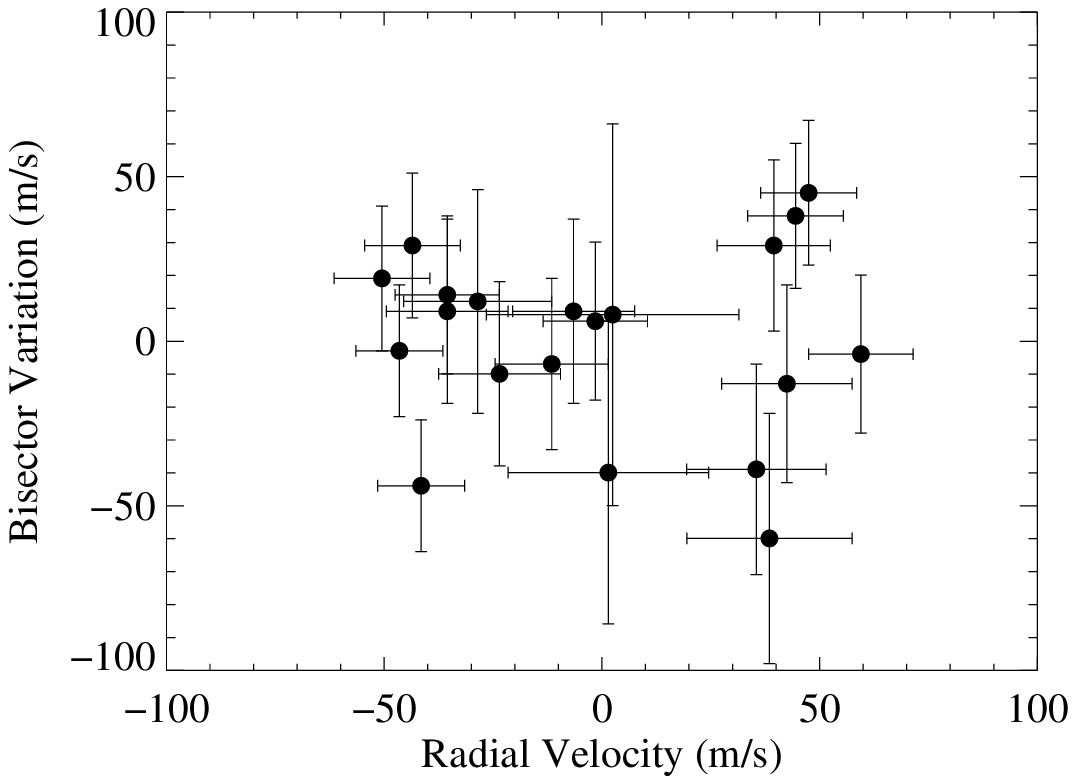}
\includegraphics[width=0.45\textwidth]{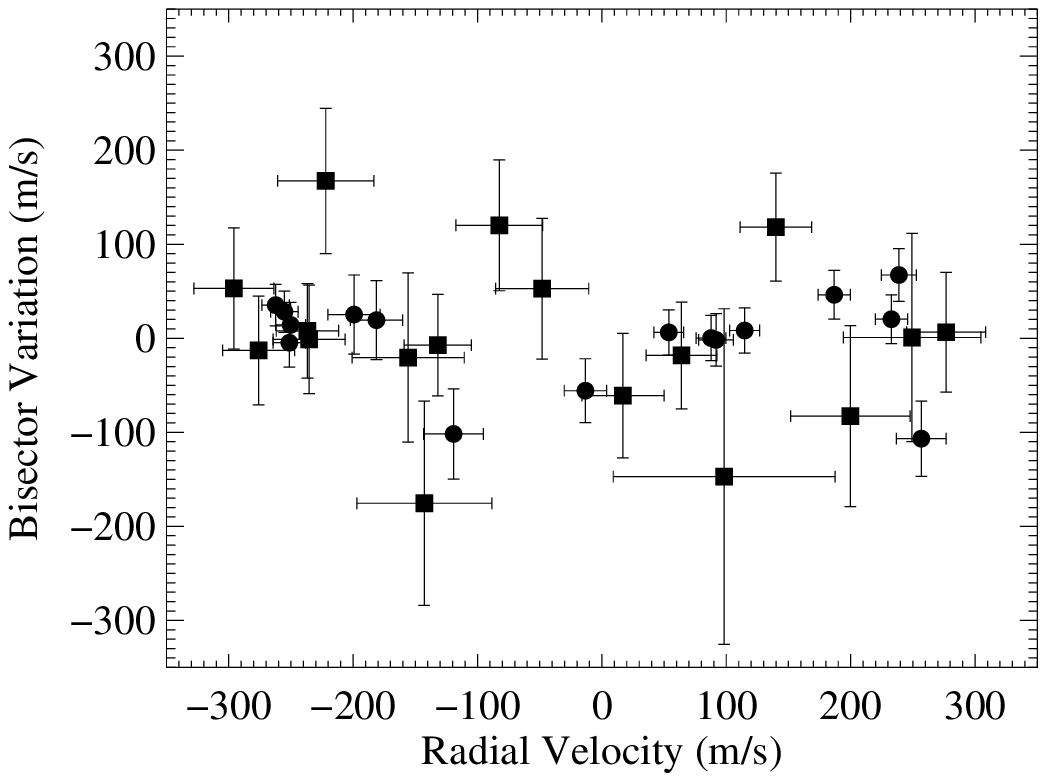}
\caption{Bisector span measurements as a function of radial velocity for WASP-113 (left) and WASP-114 (right).  {\it SOPHIE} data are plotted as circles while {\it CORALIE} are plotted as squares.\label{bissector}}
\end{figure*}

We performed a bisector span analysis for both stars that is shown in Figure~\ref{bissector}.
For both WASP-113 and WASP114 there are no significant variations of the bisector span with the radial velocities.
This supports the planetary nature of the system since in the case of stellar activity or blended eclipsing binaries the bisector can correlate with the radial velocity measurements.

\begin{table}[ht]
  \centering 
  \caption{Radial velocities of WASP-113 taken with {\it SOPHIE} at the OHP. }
  \begin{tabular}{cccc}
    \hline
    \hline
    BJD & RV & $\pm$$1\,\sigma$ & $V_{span}$ \\
    $-$2\,450\,000 & (km\,s$^{-1}$) & (km\,s$^{-1}$)  & (km\,s$^{-1}$) \\
  \hline
6403.5470&	$-$11.274&	0.014&	 0.002 \\
6472.3722&	$-$11.262&	0.014&	$-$0.017 \\
6473.4044&	$-$11.196&	0.015&	$-$0.020  \\
6475.3698&	$-$11.289&	0.011&	 0.012 \\
6476.3875&	$-$11.274&	0.012&	 0.007 \\
6477.4172&	$-$11.240&	0.012&	$-$0.001 \\
6478.3753&	$-$11.194&	0.011&	 0.031 \\
6480.3792&	$-$11.282&	0.011&	 0.022 \\
6481.4353&	$-$11.245&	0.014&	 0.002 \\
6482.4367&	$-$11.199&	0.013&	 0.022 \\
6483.3953&	$-$11.191&	0.011&	 0.038 \\
6484.3704&	$-$11.280&	0.010&	$-$0.051 \\
6485.3914&	$-$11.285&	0.010&	$-$0.010 \\
6511.3553&	$-$11.237&	0.023&	$-$0.047 \\
6514.3386&	$-$11.200&	0.019&	$-$0.067 \\
6515.3484&	$-$11.179&	0.012&	$-$0.011 \\
6516.3324&	$-$11.250&	0.013&	$-$0.014 \\
6550.3181&	$-$11.236&	0.029&	 0.001 \\
6551.3124&	$-$11.203&	0.016&	$-$0.046 \\
6552.3139&	$-$11.267&	0.017&	 0.005 \\
    \hline
  \end{tabular}
  \label{rvobservations}
\end{table}

\begin{table}[ht]
  \centering 
  \caption{Radial velocities of WASP$-$114 with {\it SOPHIE} and {\it CORALIE}.}
  \begin{tabular}{cccc}
    \hline
    \hline
    BJD & RV & $\pm$$1\,\sigma$ & $V_{span}$ \\
    -2\,450\,000 & (km\,s$^{-1}$) & (km\,s$^{-1}$)  & (km\,s$^{-1}$) \\
    \hline
      \multicolumn{3}{c}{{\it SOPHIE}     OHP} \\
    \hline
6504.6181&	$-$2.372&	0.014&	 0.024 \\
6505.6035&	$-$2.866&	0.011&	$-$0.015 \\
6507.6168&	$-$2.424&	0.013&	 0.003 \\
6508.5416&	$-$2.862&	0.013&	$-$0.048 \\
6509.4731&	$-$2.354&	0.020&	$-$0.150 \\
6510.6115&	$-$2.519&	0.014&	$-$0.045 \\
6513.6136&	$-$2.624&	0.017&	$-$0.099 \\
6514.5991&	$-$2.792&	0.021&	$-$0.024 \\
6515.6156&	$-$2.378&	0.013&	$-$0.023 \\
6516.6464&	$-$2.730&	0.024&	$-$0.145 \\
6533.4146&	$-$2.873&	0.011&	$-$0.008 \\
6534.4568&	$-$2.523&	0.012&	$-$0.043 \\
6535.3847&	$-$2.557&	0.012&	$-$0.037 \\
6536.5472&	$-$2.861&	0.012&	$-$0.029 \\
6538.5287&	$-$2.496&	0.012&	$-$0.035 \\
6550.3639&	$-$2.831&	0.029&	 ---- \\
6552.3575&	$-$2.577&	0.035&	 ---- \\
6553.4382&	$-$2.810&	0.021&	$-$0.018 \\
    \hline 
    \multicolumn{3}{c}{{\it CORALIE}     Euler} \\
    \hline
    6486.820396&             $-$3.201&              0.032&              0.034\\
    6490.770491&             $-$2.628&              0.032&             $-$0.013\\
    6505.713259&             $-$3.037&              0.027&             $-$0.026\\
    6508.752985&             $-$3.141&              0.029&             $-$0.020\\
    6511.718158&             $-$3.181&              0.029&             $-$0.032\\
    6516.657808&             $-$2.988&              0.035&              0.101\\
    6517.754717&             $-$3.127&              0.039&              0.148\\
    6531.681814&             $-$3.048&              0.054&             $-$0.195\\
    6532.612125&             $-$2.656&              0.055&             $-$0.018\\
    6533.586138&             $-$3.061&              0.045&             $-$0.040\\
    6538.604777&             $-$2.705&              0.048&             $-$0.102\\
    6540.645293&             $-$2.807&              0.089&             $-$0.167\\
    6544.625775&             $-$2.888&              0.033&             $-$0.080\\
    6545.647037&             $-$3.142&              0.025&             $-$0.011\\
    6558.604624&             $-$2.841&              0.028&             $-$0.037\\
    6561.592328&             $-$2.953&              0.037&              0.034\\
    6571.574752&             $-$2.765&              0.029&              0.099\\
    \hline
  \end{tabular}
  \label{rvobservations2}
\end{table}

\subsection{Photometric follow-up}
To better constrain the transit shape high precision transit light curves were obtained for both planets.
Seven observations were performed for WASP-113 and six for WASP-114, these are summarised in Table~\ref{obslog}.

\subsubsection{WASP-113}

Two transits of WASP-113b were observed with Near Infra-red Transiting ExoplanetS (NITES), the first on 2014 June 06 and the second on 2015 March 10. The NITES telescope is a semi-robotic $0.4$-m (f/10) Meade LX200GPS Schmidt-Cassegrain telescope installed at the observatorio del Roque de los Muchachos, La Palma. The telescope is mounted with Finger Lakes Instrumentation Proline 4710 camera, containing a $1024\times1024$ pixels deep-depleted CCD made by e2v. The telescope has a FOV and pixel scale of $11\times11$ arcmin squared and $0.66\arcsec$ pixel$^{-1}$, respectively and a peak quantum efficiency $>90\%$ at $800$ nm. For more details on the NITES Telescope we refer the reader to \citet{McCormac2014}.  For these observations, the telescope was defocused slightly to $4.0\arcsec$ FWHM and $865$ images of $20$ s exposure time were obtained with $5$ s dead time between each. Observations were obtained without a filter. The data were bias subtracted and flat field corrected using PyRAF\footnotemark \footnotetext{PyRAF is a product of the Space Telescope Science Institute, which is operated by AURA for NASA.} and the standard routines in IRAF\footnotemark \footnotetext{IRAF is distributed by the National Optical Astronomy Observatories, which are operated by the Association of Universities for Research in Astronomy, Inc., under cooperative agreement with the National Science Foundation.} and aperture photometry was performed using DAOPHOT \citep{Stetson1987}.  Two nearby comparison stars were used and an aperture radius of $6.6\arcsec$ was chosen as it returned the minimum RMS scatter in the out of transit data for WASP-113 b. Initial photometric error estimates were calculated using the electron noise from the target and the sky and the read noise within the aperture.

A transit was partially observed on 6 June 2014 at Oversky Observatory\footnote{http://www.over-sky.fr}. The telescope is a semirobotic 0.355-m ($f/10$) Celestron 14-inch Schmidt-Cassegrain installed at Nerpio (1650~m), Spain. The sequence was interrupted by clouds. The telescope is mounted with Sbig Stl-1001e camera with AOL tip tilt system, containing 1024$\times$1024 pixels (24-$\mu$m size). The telescope has a field of view of 20$\times$20 arcmin squared and a pixel scale of $1.18\sim $ arcsec/pixel. Its peak quantum efficiency is larger than 72\%\ at 775~nm. The telescope was not defocused during the acquisitions, and the exposure time was 45 seconds with 5 seconds dead time between each. Observations were obtained with a Sloan  $r'$  photometric filter. The data were dark subtracted and flat-field corrected (sky flats) using the Muniwin 2.0 software. Two nearby comparison stars and one check star were used to perform aperture phototmetry and obtain the final light curve.

Two partial transits were observed at Lowell Observatory. The first transit was obtained on 2015 June 6 with the 31" telescope. The observations had a 60 second exposure time in the V filter.
The night was not photometric and the observations ended prematurely due to clouds. The second transit was obtained on 2015 July 16 with the 42" telescope. The observations had an exposure time of 7 seconds in the V filter. Both transits were reduced with the commercial photometry package \textit{Canopus}\footnote{http://www.minorplanetobserver.com/MPOSoftware/MPOSoftware.htm}.  This package includes a photometric catalog with BVRI data derived from 2MASS JHKs photometry \citep{Warner2007}, as well as more traditional Sloan \textit{griz} and \textit{BVRI} photometry catalogs. These data provide photometric zero points and colour indices ($\sim 0.03$ mag) for the entire sky via on-chip differential photometry without the need to observe primary standards. \textit{Canopus}  returns absolute magnitudes calculated through a catalog for all stars in the image.

Another partial transit was obtained with RISE-2 mounted on the 2.3m telescope situated at Helmos observatory in Greece on 2015 July 20. The CCD size is 1k$\times$1k pixels with pixel scale of 0.51'' per pixel and a field of view of 9'$\times\,$9' \citep{Boumis2010}. The exposure time was 5 seconds and the V+R filter was used. The images were processed with IRAF for bias subtraction and flat fielding. The IRAF DAOPHOT package was used to perform aperture photometry of WASP-113 and the 7 comparison stars.

Finally a full transit was obtained with RISE at the Liverpool Telescope  on {2016 May 1}. The Liverpool telescope is a 2.0-m robotic telescope located at the Observatorio del Roque de los Muchachos on La Palma. RISE is a back illuminated, frame transfer CCD which is 1024 x 1024 pixels.  It uses a "V+R" filter and 2x2 binning of the detector for all observations. The resulting pixel scale is 1.08 arcsec/pixel.  See \citet{rise2008}. We used 2 second exposures and defocused the telescope by 1.0-mm resulting in a target FWHM of ~11.7 pix. Images are automatically bias, dark and flat corrected by the RISE pipeline. We reduced the data with standard IRAF apphot routines using a 10 pixel (10.8 arcsec) aperture.

\begin{figure}
  \centering
  \includegraphics[width=\columnwidth]{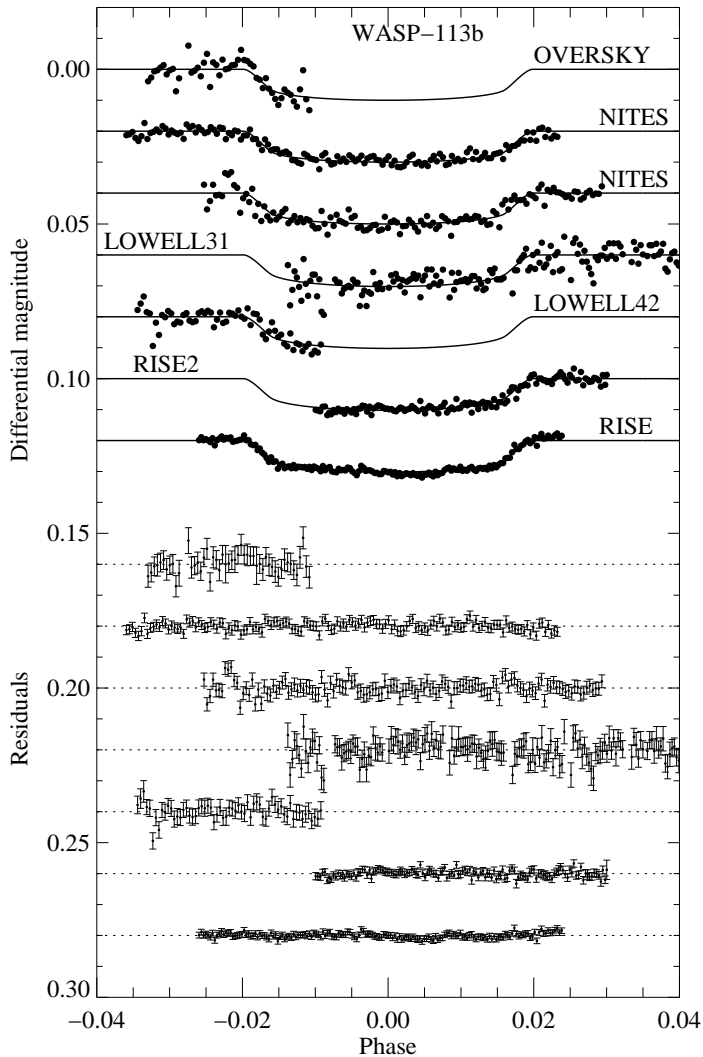}
  \caption{Phase folded light curves of WASP-113b . From top to bottom;
 Oversky taken on 2014 June 6, NITES taken on 2014 June 6 and on 2015 March 10, Lowell 31" taken on 2015 July 6, Lowell 42" taken on 2015 July 16, RISE-2 taken on 2015 July 20 and RISE taken on 2016 May 1. For each transit we superimpose the best-fit transit model. The residuals for each light curve are shown in the bottom of the figure. The data were displaced vertically and binned to 2 minutes cadence for clarity.}
  \label{photolc}
\end{figure}

The seven transit light curves of WASP-113 are shown in Figure~\ref{photolc}. We overplot the best-fit model described in detail in Section~\ref{model4}.

\subsubsection{WASP-114}

For WASP-114 six transits were obtained.  Two transits were obtained with the EulerCam, three transits were obtained with TRAPPIST and one with RISE-2.

We observed one partial and one full transit of WASP-114 using EulerCam \citep{Lendl2012} at the 1.2m Euler-Swiss telescope located at La Silla observatory. The observations were carried out on 15 September 2013 and 06 July 2014, using an I-Cousins filter and exposure times of $60$ and $90\,$s, respectively. Conditions were clear during both observations with stellar FWHMs between 1.3 and 2.4 arcsec (15 September 2013) and 1.0 and 1.5 arcsec (06 July 2014). Fluxes were extracted using a photometric aperture of 6.9 arcsec radius, and relative light curves were created using carefully selected reference stars. The final light curves have a residual RMS per 5 min bin of 580 and 490 ppm, respectively.

Three transits of WASP-114b were observed with the 0.6m TRAPPIST robotic telescope
(TRAnsiting Planets and PlanetesImals Small Telescope), located at ESO La Silla Observatory (Chile).  The transits were observed  on 29 September 2013, on19 June 2014 and on 20 July 2014. TRAPPIST is equipped with a thermoelectrically-cooled 2k$\times$2k CCD, which has a pixel scale of 0.65” that translates into a 22’$\times$22’ field of view. For details of TRAPPIST, see \citet{Gillon2011} and  \citet{Jehin2011}. The transits were observed in a blue-blocking filter\footnotemark \footnotetext{http://www.astrodon.com/products/filters/exoplanet/} that has a transmittance > 90\% from 500 nm to beyond 1000 nm. During the runs, the positions of the stars on the chip were maintained to within a few pixels thanks to a “software guiding” system that regularly derives an astrometric solution for the most recently acquired image and sends pointing corrections to the mount if needed. After a standard pre-reduction (bias, dark, and flatfield correction), the stellar fluxes were extracted from the images using the IRAF/ DAOPHOT2 aperture photometry software  \citep{Stetson1987}. For each light curve, we tested several sets of reduction parameters and kept the one giving the most precise photometry for the stars of similar brightness as the target. After a careful selection of reference stars, the transit light curves were finally obtained using differential photometry.

One more transit of WASP114b was observed with RISE-2 mounted on the 2.3m telescope situated at Helmos observatory in Greece and already described above. The images were processed with IRAF for bias subtraction and flat fielding. The IRAF DAOPHOT package was used to perform aperture photometry of WASP-114 and the six comparison stars. There is evidence for systematics in the light curve as it was also seen in the light curves taken with RISE at the Liverpool Telescope \citep{Barros2011b}. We performed the fit with and without this light curve and confirmed that the light curve does not bias the final results.

\begin{figure}
  \centering
  \includegraphics[width=\columnwidth]{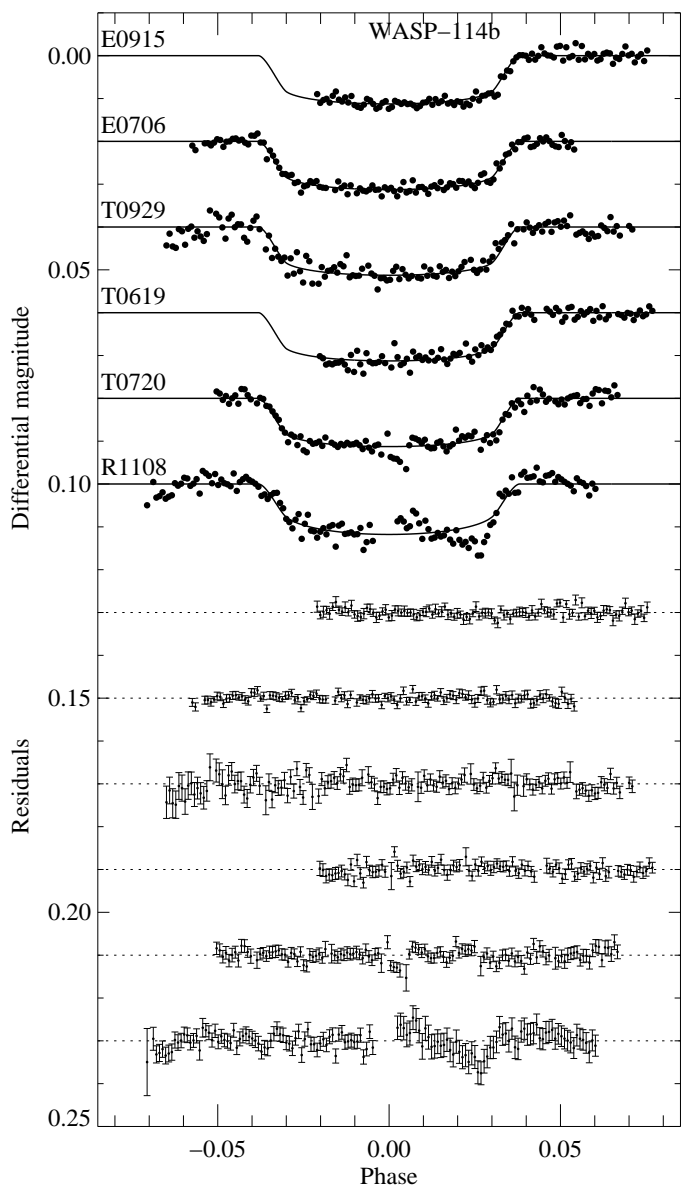}
  \caption{Phase folded light curve of WASP-114b. From top to bottom;
EulerCam observations taken on 2013 September 15 and 2014 July 6, TRAPPIST observations taken on 
 2013 September 29, 2014 June 19 and 2014 July 20 and RISE-2 observation taken on 2014 August 11.
The best-fit transit model is superimposed on the transits and the residuals of the best-fit model for each light curve are shown in the bottom of the figure. The data were displaced vertically for clarity and binned to 2 minutes cadence.}
  \label{photolc2}
\end{figure}

In Figure~\ref{photolc2} we show the six high quality transit light curves of WASP-114. The best-fit model is also overplotted.

\begin{table}[ht]
  \centering 
  \caption{Observing log for follow-up transit photometry.}
  \begin{tabular}{lcc}
    \hline
DATE & Telescope &filter \\
    \hline
  WASP-113 & & \\
    \hline
2014 June 06 & Oversky & R  \\
2014 June 06 & NITES & clear \\
2015 March 10 & NITES & clear \\
2015 July 06 &  Lowell 31" & V \\
2015 July 16 & Lowell 42" & V \\
2015 July 20 & RISE-2 & V+R \\
2016 May 01 & RISE & V+R \\
    \hline
 WASP-114& & \\
    \hline
2013 September 15& Euler & I \\ 
2013 September 29 &TRAPPIST & I+z \\
2014 June 19  &TRAPPIST & I+z \\
2014 July 6 & Euler & I \\
2014 July 20  &TRAPPIST & I+z \\
2014 August 11  &RISE2 & R \\
    \hline
  \end{tabular}
  \label{obslog}
\end{table}


\section{Spectral characterisation of the host stars}

The individual SOPHIE spectra were radial velocity corrected and co-added, giving average S/N ratios of $\sim$80:1 for both stars. The spectral analysis was performed using the procedures detailed in \cite{Doyle2013}.

For each star the effective temperature ($T_{\rm eff}$) was obtained using the H$\alpha$ line and surface gravity ($\log g$) determined from the Na~D and Mg~b lines. Iron abundances were obtained from the analysis of equivalent width measurements of several unblended Fe~{\sc i} lines. Projected rotation velocities ($v \sin I$) were determined by fitting the profiles of the Fe~{\sc i} lines after convolving with the instrumental resolution ($R$ = 75\,000) and a macroturbulent velocity adopted from the calibration of \cite{Doyle2014}. We also estimated $v \sin I$ from the CCF using the procedure of \citet{boisse2010b} and found consistent results: $5.8\pm 0.1\,$\kms\ and $4.7\pm 0.1\,$\kms\ for WASP-113 and WASP-114 respectively.

Furthermore, we estimated the spectral type using Teff  and Table B1 of \citet{Gray2008}.
The derived stellar parameters for WASP-113 and WASP-114 are given in Table~\ref{spe-params}.
We find that the stellar parameters of WASP-113 and WASP-114 are very similar.

\begin{table}[ht]
  \centering 
  \caption{Stellar parameters of WASP-113 and WASP114 from spectroscopic analysis.}
  \begin{tabular}{lll}
    \hline
Parameter & WASP-113& WASP-114\\
    \hline  
  RA(J2000)     &  14:59:29.49 & 21:50:39.74 \\
    DEC(J2000)   & +46:57:36.4 & +10:27:46.9 \\
    V [mag]       &  11.771 $\pm$ 0.045 & 12.743 $\pm$ 0.148 \\
    \teff  [K]      & 5890 $\pm$ 140 &  5940 $\pm$ 140 \\
    \logg\ [cgs] & 4.2 $\pm$ 0.1 & 4.3 $\pm$ 0.1  \\
    {[Fe/H]}  & 0.10 $\pm$ 0.09 &  0.14 $\pm$ 0.07  \\
     $v \sin I$    [ \kms ]    & 6.8 $\pm$ 0.7 & 6.4$\pm$ 0.7 \\
    log A(Li)    &   2.03 $\pm$ 0.12 & 1.77$\pm$0.12 \\
    Spectral Type &   G1 & G0 \\
    Distance  [pc]    &    360 $\pm$ 70 &460  $\pm$80 \\
    \hline
\\
  \end{tabular}
  \label{spe-params}
  \newline {\bf Note:} Spectral type estimated from Teff and Table B1 of \citep{Gray2008}.
\end{table}

\subsection{Stellar age}

\begin{figure*}
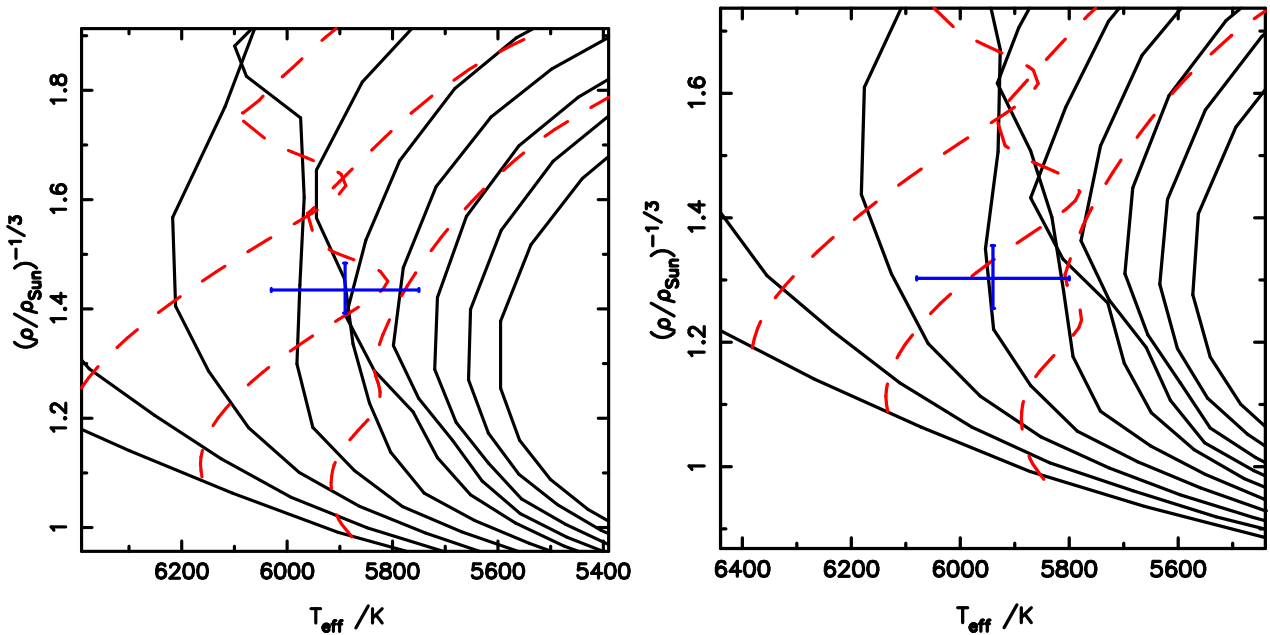

\centering
\label{iso}
\includegraphics[width=0.45\textwidth]{145929_wasp113_yonseiyFeH+0.10-1.eps}
\includegraphics[width=0.45\textwidth]{215039_wasp114_yonseiyFeH+0.14-1.eps}
\caption{Modified H-R diagrams for WASP-113 on the left and WASP-114 on the right. We display the stellar models of YY \citep{Demarque2004}. The isochrones are shown solid black line  (for $1, 2, 3, 4, 5, 6, 7, 8, 9, 10$ Gyr old stars from left to right) and and the mass tracks  in red  (from left to right we show 1.3,1.2 and 1.1 \Msun\ ) for the corresponding metallicity given in Table~\ref{spe-params}.}
\end{figure*}

Combining the \vsini\ with the stellar radius we derive an upper limit to the stellar rotation period of $10.56 \pm 1.8\,$days  and $9.96\pm 1.73\,$days for WASP-113 and WASP-114, respectively. These allow us to derive an upper limit to the gyrochronological age  of $1.04^{+0.49}_{-0.23}\,$ Gyr for WASP-113 and  $0.98^{+0.51}_{-0.31}\,$Gyr for WASP-114 using \citet{Barnes2010b}. However, the lithium abundance suggests an age of $\sim 5\,$Gyr for both stars \citep{Sestito2005}. A discrepancy between the age indicators is not unusual  \citep{Brown2014, Maxted2015}, however the reason is still not clear.  

A more precise constraint on the age can be obtained by using theoretical stellar models.
We follow the procedure of \citet{Brown2014} using three sets of stellar models: the Padova models of \citet{Marigo2008, Bressan2012}, the Yonsei-Yale (YY) isochrones of \citet{Demarque2004} and the Dartmouth Stellar Evolution Database (DSED) models of \citet{Dotter2008}.  We use the values of \teff\ and [Fe/H] derived from the spectroscopic analyses and the transit derived stellar density whose derivation is explained in the next section. 

For WASP-113 we found an age of $6.4^{+2}_{-10.9}\,$ Gyr, $5.1^{+3.3}_ {-1.7}\,$Gyr and  $7.2^{+2.9}_{-3.1}\,$Gyr, respectively for the models of Padova, YY and DSED. 
In  the left panel of Figure~\ref{iso} we show the isochrones and mass tracks for WASP-113 for the YY models. According to these models WASP-113 has a mass of $1.19^{+0.10}_{-0.13}$\Msun , which is very close to the critical mass for the stellar cores to become convective.  Due to the convective cores, the evolution and consequently the shape of the evolutionary tracks for higher mass stars is significantly different from lower mass stars as it can be seen in this figure. In this region of the parameter space details on the treatment of convective overshooting become important and the uncertainty in the stellar age and mass are higher.
We find that WASP-113 is in the end of its main sequence life and could be already in the hydrogen-shell burning phase. For WASP-114 we obtained an age of $3.9^{+2.4}_{-2.3}\,$ Gyr,  $4.2^{+1.7}_{-1.4}\,$Gyr and $4.9^{+3.1}_{-2.7}$ Gyr , respectively for the models of Padova, YY and DSED which agrees with the age estimate from the lithium abundance.  The isochrones and mass tracks for WASP-114 are shown in the right of Figure~\ref{iso}.  According to the models WASP-114 has also a mass close to the critical limit  $1.18\pm0.10$\Msun\ but the star is younger than WASP-113. WASP-114 is still in the middle of its main sequence life when the differences between the evolution of lower mass and higher mass stars are smaller.


\section{Derivation of the system parameters}

\label{model4}

To derive the system parameters we used an updated version of the Markov-Chain Monte Carlo (MCMC) fitting
procedure described by \citet{Cameron2007} and \citet{Pollacco2008}. The photometry and the radial velocity measurements are fitted simultaneously. The transit model is based on the \citet{Mandel2002} model parametrised
by the transit epoch $T_0$, orbital period $P$, impact parameter $b$,
transit duration $T_T$ and squared ratio of planet radius to star
radius $ (R_p/R_*)^2 $. For each photometric data set, we include the
non-linear limb darkening coefficients for the respective filter based
on the tables of \citet{Claret2000, Claret2004}. The  host star's reflex motion is modelled with a Keplerian parametrised by the
centre-of-mass velocity $\gamma$, the radial velocity amplitude $K$, an offset between {\it SOPHIE} and {\it CORALIE}, $\sqrt e\cos(\omega)$ and $\sqrt e\sin(\omega)$,  where $e$ is the orbital eccentricity and $\omega$ the longitude of the periastron.  The parameters $\sqrt e\cos(\omega)$ and $\sqrt e\sin(\omega)\,$ are used in order to impose a uniform prior in the eccentricity.

The \citet{Torres2010}  empirical calibration is used to estimate the stellar mass and radius from the spectroscopic derived \teff, and {[Fe/H]} and from the $\rho_*$ measured from the transit as described in \citet{Enoch2010}. In general the \citet{Torres2010} empirical calibration  was found to be in good agreement with stellar models and it has the advantage that it can be directly used in the transit fitting procedure.  For our case we also find a good agreement between the stellar mass derived using the \citet{Torres2010} empirical calibration and the one estimated using the stellar models.

The system parameters for WASP-113 and WASP-114 and the 1$\sigma$ uncertainties
derived from the MCMC analysis are given in Table~\ref{mcmc}.
For both WASP-113 and WASP-114 the transits allow a good constraint on the stellar density. Moreover, the transit derived \logg\ (for WASP113  \logg = $4.14\pm 0.05$ , for WASP114 \logg = $4.24 \pm 0.03 $ ) agrees well with the spectroscopic measured \logg\ (Table~\ref{spe-params}).  We find that despite both stars having a similar mass, WASP-113 has a lower density and larger radius confirming that WASP-113 is more evolved than WASP-114.

We find a hint of eccentricity for WASP-113 of $0.16^{+0.12}_{-0.10} $ not significant at 2 sigma. The Lucy and Sweeney test \citep{Lucy1971} gives a high probability of the eccentricity occurring by chance. Since this value of the eccentricity consistent with zero is actually non-zero it affects the transit derived stellar density changing it by slightly more than 1 sigma and biasing the system parameters. Therefore, we force a circular orbit for WASP-113b.

For WASP-114, the  eccentricity was found to be $0.01^{+0.02}_{-0.01} $, hence a circular orbit is favoured. However, in this case the eccentricity is much better constrained and the system parameters are not biased. The values of the parameters agree within 0.1 sigma for the circular and eccentric case and hence we adopt the eccentric model to obtain more reliable parameter uncertainties which are slightly bigger when the eccentricity is left free.

We conclude that WASP-113b is a $0.475\,$\Mjup\,  planet with a $1.409$ \Rjup\ radius and a low density  $0.172\, \rho_J $ in a  $4.54217\, $day orbit and WASP-114b is a  $1.769\,$\Mjup\, giant planet with a $1.339\,$ \Rjup\ radius and a density of $0.73\, \rho_J $ in a  $1.54877\,$ day orbit.

\begin{table*}[h]
  \centering 
  \caption{WASP-113 and WASP-114 system parameters.}
  \label{mcmc}
  \begin{tabular}{lcccl}
    \hline
    \hline
    Parameter & WASP-113 & WASP-114 \\
    \hline
    Transit epoch $T_0$ [HJD] & $ 2457197.097459  \pm 0.000 040 $ & $ 2456667.73582  \pm 0.00021 $ \\
    Orbital period $P$ [days] & $ 4.54216874538\pm 0.0000042 $ &$ 1.5487743^{+0.000 001 2}_{-0.00000091} $ \\                                                 
    Planet/star area ratio $ (R_p/R_*)^2 $ &  $0.00809 \pm 0.00028$ &  $0.00927 \pm 0.00016$ \\
    Transit duration $T_T$ [days] & $  0.1791 \pm 0.0019$ &  $ 0.11592 \pm 0.00079$  \\
    Impact parameter $ b $ [$R_*$]  & $ 0.486^{+0.063}_{-0.14}$ &  $ 0.457^{+0.052}_{-0.054}$ \\
    Orbital inclination $ I $ [degrees] & $ 86.46^{+1.2}_{-0.64}$ & $ 83.96 \pm 0.90$ \\
    &    &      &  \\
    Stellar reflex velocity $K$ [\ms] & $ 48.74\pm 5.2 $ & $ 260.6 \pm 5.4 $ \\
    Orbital semimajor axis $ a $ [au] & $ 0.05885 \pm 0.00010 $ &  $0.02851 \pm   0.00039$  \\
    Orbital eccentricity  $ e $ &  fixed=0  &    $0.012^{+0.022}_{-0.009}$ \\
    Longitude of periastron  $ \omega $[\degree] & fixed=0 & $ -71.^{+150}_{-35}$ \\
    &    &      &  \\
    Stellar mass $ M_* $ [\Msun] & $   1.318 \pm 0.069$  & $   1.289 \pm 0.053$ \\
    Stellar radius $ R_* $ [\Rsun]  & $ 1.608 ^{+ 0.090}_{-0.12}  $  & $ 1.43\pm {0.060}  $  \\
    Stellar density $ \rho_* $ [$\rho_\odot$] & $ 0.317^{0.072}_{-0.039}  $   & $0.439^{+0.050}_{-0.041}$ \\
    &    &      &  \\
    Planet mass  $ M_p $ [\Mjup]  & $  0.475^{+0.054}_{-0.052}$ & $ 1.769 \pm 0.064$ \\
    Planet radius $ R_p $ [\Rjup]  & $ 1.409 ^{+ 0.096}_{-0.14} $  & $ 1.339 \pm 0.064 $  \\
    Planet density  $ \rho_p $ [$\rho_J$]  & $ 0.172^{+0.055}_{-0.034}$   & $ 0.73  \pm 0.10 $  \\
Planet surface gravity  $\mbox{log}g_p $  [cgs] & $ 2.744^{+0.081}_{-0.072} $ &  $3.353 \pm 0.036 $\\ 
Planet temperature $T_{eq}$ [K]  & $ 1496 \pm 60 $& $2043 \pm 58 $\\ 
    \hline
  \end{tabular}

\end{table*}

\section{Discussion}

We present the discovery and characterisation of WASP-113b, a hot Jupiter in a  $4.5\, $day orbit around a G1 type star, and of WASP-114b, a hot Jupiter in a 
$1.5\,$ day orbit around a G0 star. WASP-113, with $\teff=5890\,$K,  \logg$=4.2$, and [Fe/H]$=0.10$, is orbited by a $0.475\,$\Mjup\,  planet with a radius of $1.409$ \Rjup\ and hence a density of  $0.172\, \rho_J $ . WASP-114 has  a $\teff=5940\,$K,  \logg$=4.3$, and [Fe/H]$=0.14$,  and the planetary companion is more massive with a mass of $1.769\,$\Mjup\,  and a radius of $1.339\,$ \Rjup\  and hence a density of $0.73\, \rho_J $.

A circular orbit is preferred for both planets which is not surprising given that the circularisation timescale \citep{Goldreich1966,Bodenheimer2001} is short compared to the age of the host stars.  Using equation 2 of \citet{Bodenheimer2001} and assuming $Q_p=10^5-10^6$ \citep{Levrard2009},  we compute the circularisation timescale for WASP-113b,  $\tau_{cir}= 0.016-0.16\,$ Gyr and for WASP-114b, $\tau_{cir}= 0.00073-0.0073\,$Gyr.  These are much shorter than the host stars ages derived from spectra ( $\sim 6.2\,$Gyr and  $\sim 4.3\,$Gyr  for WASP-113 and WASP-114 respectively), thus favouring circular orbits.

The mass-radius relationship for giant planets depends on their internal composition, since heavier elements decrease the planetary radius. Assuming the coreless models of \citet{Fortney2007} we estimated a radius of 1.05 \Rjup\ for WASP-113b and 1.15 \Rjup\ for WASP-114b. Both these predicted radii are more than $2\sigma$ smaller than the radii measured for the planets in our analysis. Hence, we conclude that the planets are inflated. Following \citet{Laughlin2011} we compute a radius anomaly, $\Re=0.35$, for WASP-113b and  $\Re=0.189$ for WASP-114b. Recently it has become clear that regardless of the nature of the inflation mechanism there is a clear correlation between the stellar incident flux and the radius anomaly \citep{Laughlin2011, Weiss2013}. Furthermore, giant planets that receive modest stellar flux do not show a radius anomaly \citep{Demory2011}. 
At first glance the radius anomaly of WASP-113b and WASP-114b seem to contradict this correlation since the effective temperature of WASP-114b is higher than WASP-113b.
In fact WASP-113b has radius anomaly above the mean relationship proposed by \citet{Laughlin2011} ( $\Re \propto T_{equ}^{1.4}$ ) while WASP-114b has a radius anomaly slightly lower than this mean scaling relationship.  However, other exoplanets share the same properties as WASP-113b and WASP-114b and the differences of radius anomaly can be explained by the planetary mass. WASP-114 is 3.7 times more massive than WASP-113. Therefore, it has probably a much higher amount of planetary heavy elements and its higher gravitational binding energy contra-acts the inflation.  The known exoplanets in the mass range of  0.5\Msun\ and 1.5\Msun\ have the largest known radii\citep{Lopez2016} in agreement the higher radius anomaly of WASP-113b.

There are several theories to explain the inflated radii in hot Jupiters: tidal heating \citep{Bodenheimer2001}; enhanced atmospheric opacities
\citep{Burrows2007};  kinetic heating
due to winds \citep{Guillot2002} and ohmic dissipation
\citep{Batygin2011}. 
The two last are mechanisms related to incident stellar flux and are favoured by the recent reported correlation. However, there is no theory capable of explaining all the measured radius anomalies  \cite[e.g.][]{Spiegel2013}.  Due to the old age of both WASP-113 and WASP-114 compared with the circularisation timescale it is not expected that tidal heating would be playing an important role hence a combination of the other mechanisms is more likely.

WASP-113 is at the end of its main sequence life. We estimate that due the expansion of the star after the end of the main sequence the planet with be engulfed in $\sim1.4$Gyr. The response of the planetary radius to the increase of the stellar radius and stellar luminosity will distinguish mechanisms that are directly capable of inflating the planet from those that only slow down the cooling exoplanets \citep{Lopez2016}. An alternative to this long wait is to search for hot-Jupiters with periods 10-30 days around evolved stars \citep{Lopez2016}.

Follow-up observations of these planets can help shed light on the radius anomaly and on the atmospheric composition of these planets. The large scale-height of WASP-113b, $\sim 950$ km, and its relatively bright host star, V = 11.8 , makes it a good target for transmission spectroscopy observations to probe its atmospheric composition. The scale height of  WASP-114b is smaller  $\sim 310$ km and combined with its fainter host star would make atmospheric studies more challenging.


\begin{acknowledgements}
The WASP Consortium consists of astronomers primarily from Queen's
  University Belfast, St Andrews, Keele, Leicester, The Open
  University, Isaac Newton Group La Palma and Instituto de Astrofsica
  de Canarias. The SuperWASP-N camera is hosted by the Issac Newton
  Group on La Palma. We are grateful for their support and
  assistance. Funding for WASP comes from consortium universities and
  from the UK’s Science and Technology Facilities Council.   
 Based on observations
  made at Observatoire de Haute Provence (CNRS), France and at the ESO
  La Silla Observatory (Chile) with the {\it CORALIE} Echelle
  spectrograph mounted on the Swiss telescope. We thank the staff at Haute-Provence Observatory.
 SCCB acknowledges support by grant  98761 from CNES and  the Funda\c c\~ao para a Ci\^encia e a Tecnologia  (FCT) through the Investigador FCT Contract No. IF/01312/2014 and the grant reference PTDC/FIS-AST/1526/2014. 
 D.J.A., D.P. and C.W. acknowledge funding from the European Union Seventh Framework programme (FP7/2007- 2013) under grant agreement No. 313014 (ETAEARTH).
 The Swiss {\it Euler} Telescope is operated by the University of Geneva, and is funded by the Swiss National Science Foundation. The Aristarchos telescope is operated on Helmos Observatory by the Institute for Astronomy, Astrophysics, Space Applications and Remote Sensing of the National Observatory of Athens.
TRAPPIST is a project funded by the Belgian Fund for Scientific Research (Fond National de la Recherche Scientifique, F.R.S-FNRS) under grant FRFC 2.5.594.09.F, with the participation of the Swiss National Science Fundation (SNF).  M. Gillon and E. Jehin are FNRS Research Associates.  L. Delrez acknowledges support of the F.R.I.A. fund of the FNRS. 
\end{acknowledgements}

\bibliographystyle{aa}
 \bibliography{susana}

\end{document}